# Thermodynamics of quasianti-Hermitian quaternionic systems

S. A. Alavi

Department of Physics, Hakim Sabzevari university , P. O. Box 397, Sabzevar, Iran.

s.alavi@hsu.ac.ir ; alialavi@fastmail.us

**Abstract**: *Thermodynamics of quasianti-Hermitian quaternionic systems with constant number of particles in equilibrium is studied. A toy model is introduced and the physically relevant quantities are derived. The energy fluctuation which shows that for large N the relative r.m.s fluctuation in the values of E is quite negligible is derived. The negative temperature for such systems is also studied. Finally a physical example is discussed and physical explanations in the context of quantum physics are given.*

**1-Introduction**

Non-Hermitian Hamiltonians are currently an active field of research, motivated by the necessity to understand the mathematical properties of their subclasses, namely the pseudo-Hermitian and pseudoanti-Hermitian Hamiltonians. From physical point of view the investigation of existence of a suitable similarity transformation that maps such Hamiltonians to an equivalent Hermitian form has caused much concern. Also there have been efforts to look for non-Hermitian Hamiltonians which have a real spectrum such that the accompanying dynamics is unitary.

The interest in non-Hermitian Hamiltonians was put forward by Bender and Boettcher [1]. They pointed out that PT-symmetric Hamiltonians could possess real bound-state eigenvalues. Then Mostafazadeh [2] shown that the reality of the spectrum is ensured [3] if the Hamiltonian H is Hermitian with respect to a positive-definite inner product $< . , .>_+$ on the Hilbert space H in which H is acting [4].

A Hamiltonian H is called pseudo-Hermitian if it satisfies the relation $H^\dagger = \eta H \eta^{-1}$, where $\eta$ is a linear, Hermitian and invertible operator. One can also express this definition in the form $H^\circ = H$, where $H^\circ = \eta^{-1} H^\dagger \eta$ is the $\eta$-pseudoadjoint of H. Unlike Hermiticity which is a sufficient condition for the reality of the spectrum, pseudo-Hermiticity is a necessary condition. To have a condition which is both necessary and sufficient, the metric operator must be positive- definite that is it must be of the form $\eta_+ = \Theta^\dagger \Theta$ [2]. Such a positive-definite operator defines a positive-definite inner product $<.,.>_+ := <.|\eta.>$, where $<.,.>$ is the defining inner product of the Hilbert space in which H and $\eta_+$ act.

The foundations of quaternionic quantum mechanics(QQM) were laid by Finkelstein et.al., in the 1960's[5]. A systematic study of QQM is given in [6]. Moreover, the theory of open quantum systems can be obtained, in many relevant physical situations, as the complex projection of quaternionic closed quantum systems [7].

Experimental tests on QQM were proposed by Peres [8] and carried out by Kaiser et.al.,[9] searching for quaternionic effects manifested through non-commuting scattering phases when a particle crosses a pair of potential barriers. See also [10]. A review of the experimental status of QQM can be found in [11].

On the other hand theoretical framework of CQM(complex quantum mechanics) has been extended and generalized by introducing the pseudo-Hermitian operators. By the same motivation if one wishes to extend and generalize the theoretical framework of standard quaternionic Hamiltonians and symmetry generators, one needs to introduce and study the pseudoanti-Hermitian quaternionic operators. This motivates us to study the physics of quasianti-Hermitian quaternionic systems. In [12] the author generalized the framework of statistical mechanics and thermodynamics to pseudo-Hermitian operators. In this paper the framework of statistical mechanics and thermodynamics is generalized to pseudoanti-Hermitian quaternionic systems.

It is worth mentioning that while in both QQM and CQM theories, observables are associated with self-adjoint (or Hermitian) operators, the Hamiltonians are Hermitian in CQM, but they are anti-Hermitian in QQM, and the same happens for the symmetry generators, like the angular momentum operators.

A quaternion is usually expressed as $q = q_0 + iq_1 + jq_2 + kq_3$ where $q_{0,1,2,3} \in R$ and $i^2 = j^2 = k^2 = ijk = -1$, $ij = -ji = k$, and with an involutive anti-automorphism(conjugation) such that, $q \to \bar{q} = q_0 - iq_1 - jq_2 - kq_3$.

By a definition a quaternionic linear operator H is said to be ($\eta_-$)pseudoanti-Hermitian if a linear invertible Hermitian operator $\eta$ exists such that $\eta H \eta^{-1} = -H^\dagger$. If $\eta$ is positive definite, H is said quasianti-Hermitian.

The density matrix $\rho_\psi$ associated with a pure state belonging to a quarternionic n-dimensional Hilbert space $Q^n$ is defined by : $\rho_\psi \equiv \rho = |\psi\rangle\langle\psi|$

In QQM the dynamics of the quantum system is described by the following Schroedinger equation:

$$\frac{d}{dt}|\psi\rangle = -H|\psi\rangle \tag{1}$$

It is easy to show that the density matrix satisfies in the same equation $\frac{\partial \rho}{\partial t}$ =-H$\rho$.

Let us now consider the space of quaternionic quasi-Hermitian density matrices ,that is the subclass of $\eta$ −pseudo-Hermitian density matrices with a positive $\eta$.Thus, an (Hermitian) operator $\theta$ exists such that $\eta = \theta^2$. Then the inner product $(.,.)_\eta = (.,\eta.)$ introduced in the Hilbert space is positive which is necessity for a quantum measurement.

The most general 2-dimensional complex positive $\eta$ operator is given by:

$$\eta = \theta^2 = \begin{pmatrix} x & z \\ z^* & y \end{pmatrix}\begin{pmatrix} x & z \\ z^* & y \end{pmatrix} = \begin{pmatrix} x^2 + |z|^2 & (x+y)z \\ (x+y)z^* & y^2 + |z|^2 \end{pmatrix}$$

where $x, y \in R$; $z \in C$ and $x, y \neq |z|^2$[13]. The choice of complex metric operators can be justifed as follows: it is shown in [13] that the complex projections of time-dependent $\eta$-quasianti-Hermitian quaternionic Hamiltonian dynamics are complex stochastic dynamics in the space of complex quasi-Hermitian density matrices if and only if a quasi stationarity condition is fulfilled, i.e., if and only if η is an Hermitian positive time-independent complex operator.

## 2- Quasianti-Hermitian quaternionic systems.

Now let us assume that H describes an ensemble of a huge but fixed number of independent particles. In what follows we study the statistical description of the system in equilibrium. The state of the system is characterized by a density matrix ρ which can be written in energy representation as :

$$\rho = \sum_n w_n |n\rangle\langle n| \quad , \mathrm{Tr}\rho < \infty \tag{2}$$

In equilibrium, the density matrix operator is solution of the following equation :

$$\frac{\partial \rho}{\partial \beta} = -H\rho \tag{3}$$

where $\beta = \frac{1}{kT}$ and k is the familiar Boltzmann constant. In the canonical ensemble the macrostate of a member system is defined through the parameters N, V, T and the enegry E is a variable quantity. The density matrix in the energy representation has the following form $\hat{\rho} = \frac{e^{-\beta \hat{H}}}{Z}$ where $Z = \mathrm{Tr}(e^{-\beta \hat{H}})$ is the partition function. The partition function plays an important role in thermodynamics of the system because it allows direct computation of thermodynamic quantities.

On the other hand Eq.(3) resembles Schrodinger equation for evolution operator in QQM. The link can be established by the substitution $\beta \to t$. Working in quasianti-Hermitian picture the Eq.(3) reads:

$$\frac{\partial U}{\partial t} = -HU \; ; \quad U(0) = 1 \tag{4}$$

where we defined $U(t) = \rho(t)$. In the x-representation, we have

$$\rho(x_1, x_2)_{\beta \to t} = U(x_1, x_2; t) = \langle x_1|e^{-Ht}|x_2\rangle \equiv : G(x_1, x_2; t). \tag{5}$$

this propagator is of fundamental importance because it allows to compute partition function $(Z = \int G(x_1, x_2; -\beta)dx)$. The analogous equation in the case of pseudo-Hermitian Hamiltonian has been derived in [12].

## 3- Statistical mechanics of quasianti-Hermitian quaternionic systems. A Toy model.

Let's consider an ensemble of systems each consisting of N distinguishable particles without mutual interaction. The partition function can be written as follows:

$$Z = (Z_1)^N \tag{6}$$

where $z_1$ may be regarded as the partition function of a single particle in the system. In the quasianti-Hermitian quaternionic picture, the subsystem is described by the following most general Hamiltonian:

$$H = \begin{pmatrix} a & c \\ d & b \end{pmatrix}$$

Here a, b and c are three arbitrary quaternion. The requirement that the Hamiltonian be quasianti-Hermitian quaternionic with respect to $\eta$ i.e. $\eta H \eta^{-1} = -H^\dagger$ gives: $a^* = -a$, $b^* = -b$ and $d = -\frac{\alpha}{\gamma} c^*$ where $\eta$ is given by :

$$\eta = \theta^2 = \begin{pmatrix} \alpha & 0 \\ 0 & \gamma \end{pmatrix}$$

Here z, x and y have been chosen as follows : $z = 0$, $x^2 = \alpha$ and $y^2 = \gamma$. $\alpha$ and $\gamma$ are two arbitrary positive real numbers. The energy levels of the system may depend on the external parameters e.g. on the volume in which the particle is confined.
The partition function can be found by solving Eq.(4). By separating the Hamiltonian into its diagonal and non-diagonal parts, we have

$$H = H_0 + H' \tag{7}$$

where :

$$H_0 = \begin{pmatrix} a & 0 \\ 0 & b \end{pmatrix}, \quad H' = \begin{pmatrix} 0 & c \\ -\frac{\alpha}{\gamma} c^* & 0 \end{pmatrix}$$

Let us introduce the following transformation :

$$U_0(t) = \exp(-H_0 t) \tag{8}$$

A new operator in the interaction picture is defined as follows:

$$U_I(t) = U_0^\dagger(t) U(t) U_0(t) \tag{9}$$

Then Eq.(4) changes into the following form :

$$\frac{\partial U_I}{\partial t} = -H_I'(t) U_I(t); \quad U_I(0) = 1 \tag{10}$$

where $H'(t) = U_0(t)^\dagger H' U_0(t)$. One can find the solution iteratively by integrating both sides of the equation.

$$U_I(t) = 1 - \int_0^t dt' H_I(t') + \int_0^t dt' \int_0^{t'} dt'' H_I'(t') H_I'(t'') + \cdots \tag{11}$$

where the higher order terms has been neglected, then :

$$U_I(t) \sim \begin{pmatrix} 1 + \frac{\alpha}{\gamma}\frac{c^2}{a-b}t - \frac{\alpha}{\gamma}\frac{c^2}{(a-b)^2}(e^{(a-b)t} - 1) & \frac{c}{b-a}(e^{(a-b)t} - 1) \\ -\frac{\gamma}{\alpha}\frac{d}{a-b}(e^{-(a-b)t} - 1) & 1 - \frac{\alpha}{\gamma}\frac{c^2}{a-b}t - \frac{\alpha}{\gamma}\frac{c^2}{(a-b)^2}(e^{-(a-b)t} - 1) \end{pmatrix}$$

The approximate value of the partition function is as follows :

$$Z_1(\beta) = Tr(U_0(t)U_I(t))\big|_{t \to \beta} \sim e^{-a\beta} + e^{-b\beta} - \frac{\alpha}{\gamma}\frac{c^2}{a-b}\beta(e^{-b\beta} - e^{-a\beta}) \tag{12}$$

Now the calculation of the thermodynamic quantities is straightforward. For the Helmholtz free energy of the system we have :

$$A = -NkT \ln Z_1$$

$$= -\frac{N}{\beta}\ln Z_1 = -\frac{N}{\beta}\ln[e^{-a\beta} + e^{-b\beta} - \frac{\alpha}{\gamma}\frac{c^2}{a-b}\beta(e^{-b\beta} - e^{-a\beta})] \tag{13}$$

The entropy of the system is :

$$S = Nk \ln\left[e^{-a\beta} + e^{-b\beta} - \frac{\alpha}{\gamma}\frac{c^2}{a-b}\beta(e^{-b\beta} - e^{-a\beta})\right]$$

$$+ Nk\beta \frac{\left[a(a-b) - \frac{\alpha}{\gamma}c^2 + \frac{\alpha}{\gamma}c^2\beta a\right]e^{\beta b} + \left[b(a-b) + \frac{\alpha}{\gamma}c^2 - \frac{\alpha}{\gamma}c^2\beta b\right]e^{\beta a}}{\left[(a-b) + \frac{\alpha}{\gamma}c^2\beta\right]e^{\beta b} + \left[(a-b) - \frac{\alpha}{\gamma}c^2\beta\right]e^{\beta a}} \tag{14}$$

The internal energy of the system is given by :

$$U = N \frac{e^{\beta b}\left[a\left(a-b+\frac{\alpha}{\gamma}c^2\beta\right) - \frac{\alpha}{\gamma}c^2\right] + e^{\beta a}\left[b\left(a-b-\frac{\alpha}{\gamma}c^2\beta\right) + \frac{\alpha}{\gamma}c^2\right]}{e^{\beta b}\left[a-b+\frac{\alpha}{\gamma}c^2\beta\right] + e^{\beta a}\left[a-b-\frac{\alpha}{\gamma}c^2\beta\right]} \tag{15}$$

The specific heat $C_V$ which describes how the temperature changes when the heat is absorbed while volume V of the system remains unchanged is given by :

$$C_V = \left(\frac{\partial U}{\partial T}\right)_V = -\beta^2 k\left(\frac{\partial U}{\partial \beta}\right)$$

$$= \beta^2 kN \frac{e^{\beta(a+b)}\left\{(a-b)^2\left[(a-b)^2 - \left(\frac{\alpha}{\gamma}c^2\beta\right)^2 - 4\frac{\alpha}{\gamma}c^2\right] + 2\left(\frac{\alpha}{\gamma}c^2\right)^2\right\} - \left(\frac{\alpha}{\gamma}c^2\right)^2[e^{2\beta a} + e^{2\beta b}]}{\{e^{\beta b}\left[a-b+\frac{\alpha}{\gamma}c^2\beta\right] + e^{\beta a}[a-b-\frac{\alpha}{\gamma}c^2\beta]\}^2}$$

(16)

The pressure of the system is as follows :

$$P = -\left(\frac{\partial A}{\partial V}\right)_\beta = N \frac{\left[-(a-b)^2 a' - \frac{\alpha}{\gamma} c^2 (\beta(a-b) + 1)a' + \frac{\alpha}{\gamma} c^2 b' + 2\frac{\alpha}{\gamma} c(a-b)c'\right]}{\left[(a-b)^2 + \frac{\alpha}{\gamma} c^2 \beta(a-b)\right] e^{-a\beta} + \left[(a-b)^2 - \frac{\alpha}{\gamma} c^2 \beta(a-b)\right] e^{-b\beta}} e^{-a\beta} +$$

$$N \frac{\left[-(a-b)^2 b' - \frac{\alpha}{\gamma} c^2 (1-\beta(a-b))b' + \frac{\alpha}{\gamma} c^2 a' - 2\frac{\alpha}{\gamma} c(a-b)c'\right]}{\left[(a-b)^2 + \frac{\alpha}{\gamma} c^2 \beta(a-b)\right] e^{-a\beta} + \left[(a-b)^2 - \frac{\alpha}{\gamma} c^2 \beta(a-b)\right] e^{-b\beta}} e^{-b\beta} \tag{17}$$

where the prime symbol represents the derivative with respect to the volume, $a' = \frac{\partial a}{\partial V}$.

## 4- Energy fluctuation.

We first write down the expression for the mean energy :

$$U = <E> = \frac{\sum_r E_r g_r \exp(-\beta E_r)}{\sum_r g_r \exp(-\beta E_r)} \tag{18}$$

where $g_r$ is the multiplicity of a particular energy level $E_r$. By differentiating of the expression of the mean energy with respect to the parameter $\beta$, we obtain :

$$\frac{\partial U}{\partial \beta} = -\{\langle E^2 \rangle - \langle E \rangle^2\} \tag{19}$$

whence it follows that :

$$\langle (E)^2 \rangle = \langle E^2 \rangle - \langle E \rangle^2 = -\frac{\partial U}{\partial \beta} = kT^2 \left(\frac{\partial U}{\partial T}\right) = kT^2 C_v \tag{20}$$

So, for the relative root-mean-square fluctuation in E, we have :

$$\frac{\sqrt{\langle (E)^2 \rangle}}{\langle E \rangle} = \frac{\sqrt{kT^2 C_v}}{U} = \frac{1}{\sqrt{N}} \frac{\beta \sqrt{k\, f(a,b,c,\alpha,\beta,\gamma)}}{g(a,b,c,\alpha,\beta,\gamma)} \propto \frac{1}{\sqrt{N}} \tag{21}$$

where $f(a, b, c, \alpha, \beta, \gamma)$ is the numerator of Eq.(16) and $g(a, b, c, \alpha, \beta, \gamma)$ is the numerator of Eq.(15). As we observe it is $O(N^{-\frac{1}{2}})$, N being the number of particles in the system. Consequently, for large N(which is true for every statistical system), the relative r.m.s fluctuation in the values of E is quite negligible. Thus for all practical purposes, a system in the canonical ensemble in quasianti-Hermitian quaternionic picture, has an energy equal to or almost equal to the mean energy U; the situation is therefore practically the same as in the microcanonical ensemble.

## 5- Negative temperature.

Let us study the following problem : in how many different ways, can our system attain a state of energy E?. This can be tackled in precisely the same way as the problem of the random walk. Let $N_+$ be the number of particles with energy $E_+$ and $N_-$ with energy $E_-$; then

$$E = E_+ N_+ + E_- N_- \quad , \quad N = N_+ + N_- \tag{22}$$

Solving for $N_+$ and $N_-$ we obtain :

$$N_+ = \frac{E - NE_-}{E_+ - E_-}, \quad N_- = \frac{E - NE_+}{E_- - E_+} \tag{23}$$

The desired number of ways is then given by the expression:

$$\Omega(E, N) = \frac{N!}{N_+! N_-!} = \frac{N!}{\left(\frac{E-NE_-}{E_+-E_-}\right)! \left(\frac{E-NE_+}{E_--E_+}\right)!} \tag{24}$$

whence we obtain for the entropy of the system :

$$S(N, E) = \kappa \ln \Omega \simeq$$

$$\kappa \left[ N \ln N - \frac{E - NE_+}{E_- - E_+} \ln \left( \frac{E - NE_+}{E_- - E_+} \right) - \frac{E - NE_-}{E_+ - E_-} \ln \left( \frac{E - NE_-}{E_+ - E_-} \right) \right] \tag{25}$$

The temperature of the system is as follows:

$$\frac{1}{T} = \left( \frac{\partial S}{\partial E} \right)_N = \frac{\kappa}{E_- - E_+} \ln \left( -\frac{E - NE_-}{E - NE_+} \right) \tag{26}$$

One can see from Eq.(26) that, so long as $E < \frac{N}{2}(E_+ + E_-)$, $T > 0$. However the same equation tells us that if $E > \frac{N}{2}(E_+ + E_-)$, then $T < 0$. Lets examine the matter a little more closely. For this purpose we consider as well the variation of the entropy S with the energy E. We note that for $E = NE_-$, both S and T vanish. As E increase, they too increase until we reach the special situation where $E = \frac{N}{2}(E_+ + E_-)$. The entropy is then seen to have attained its maximum value $S = Nk \ln 2$, while the temperature has reached an infinite value. Throughout this range, the entropy was a monotonically increasing function of energy, so T was positive. Now as E equals $\left[\frac{N}{2}(E_+ + E_-)\right]_+$, $\left(\frac{dS}{dE}\right)$ becomes $0_-$ and T becomes $-\infty$. With a further increase in the value of E, the entropy monotonically decreases; as a result, the temperature continues to be negative, though its magnitude steading decreases. Finally, we reach the largest value of E, namely $NE_+$, where the entropy is once again zero and $T = -0$.

**6- Physical example : A spin one half system in a constant quasianti-Hermitian quaternionic potential.**

Let us consider a two-level quantum system with a quasianti-Hermitian quaternionic Hamiltonian $H = H_\alpha + j H_\beta$. $H_\alpha$ denotes the free complex anti-Hermitian Hamiltonian describing a spin half particle in a constant magnetic field

$$H_\alpha = \frac{w}{2}\begin{pmatrix} i & 0 \\ 0 & -i \end{pmatrix}$$

and $jH_\beta$ is a purely quasianti-Hermitian quaternionic constant potential :

$$jH_\beta = \begin{pmatrix} 0 & j\frac{v}{x} \\ jvx & 0 \end{pmatrix}; \quad v, x \neq 0 \in R.$$

$H = H_\alpha + jH_\beta$ is $\eta'$-quasianti-Hermitian quaternionic i.e.,
$\eta' H \eta'^{-1} = -H^\dagger$, where :

$$\eta' = \begin{pmatrix} x^2 & 0 \\ 0 & 1 \end{pmatrix}$$

The eigenvalues and the corresponding biorthonormal eigenbasis of the quaternionic Hamiltonian H are [14]:

$$iE_\pm = i\left(\frac{w}{2} \pm v\right) \tag{27}$$

and:

$$|\psi_\pm\rangle = \begin{pmatrix} \pm\frac{i}{x} \\ j \end{pmatrix}\frac{1}{\sqrt{2}}, \quad |\phi_\pm\rangle = \begin{pmatrix} \pm xi \\ j \end{pmatrix}\frac{1}{\sqrt{2}}$$

It is worth mentioning that the metric $\eta'$ is a special case of the metric proposed in section 2 and can be obtained by substituting $\alpha = x^2$ and $\gamma = 1$. The Hamiltonian H is also a special case of the general Hamiltonian introduced in section 2 and can be obtained by substituting the values of a, b, c and d. So the thermodynamic properties of this system can be obtained straightforwardly from equations (13-17), for instance we have :

$$S = N\kappa \ln\left[2\left(\cosh\frac{w\beta}{2} - \frac{v^2\beta}{w}\sinh\frac{w\beta}{2}\right)\right]$$

$$+N\kappa\beta\frac{-w^2\sinh\frac{w\beta}{2} + 2v^2\sinh\frac{w\beta}{2} + \beta wv^2\cosh\frac{w\beta}{2}}{2w\cosh\frac{w\beta}{2} - 2\beta v^2\sinh\frac{w\beta}{2}} \tag{28}$$

$$U = N\frac{-w^2\sinh\frac{w\beta}{2} + 2v^2\sinh\frac{w\beta}{2} + \beta wv^2\cosh\frac{w\beta}{2}}{2w\cosh\frac{w\beta}{2} - 2\beta v^2\sinh\frac{w\beta}{2}} \tag{29}$$

Moreover the negative temperature discussed in previous section is applicable to this system which is studied now. Let $N_+$ be the number of particles with energy $E_+ = \frac{w}{2} + v$ and $N_-$ with energy $E_- = \frac{w}{2} - v$, then we have :

$$E = E_+N_+ + E_-N_- \quad , \quad N = N_+ + N_- \tag{30}$$

The number of ways this system can attain a state of energy E is given by:

$$\Omega(E, N) = \frac{N!}{N_+!N_-!} = \frac{N!}{\left(-\frac{E-N(\frac{w}{2}+v)}{2v}\right)!\left(\frac{E-N(\frac{w}{2}-v)}{2v}\right)!} \tag{31}$$

So the entropy of the system is as follows:

$$S(N, E) = \kappa \ln \Omega \simeq \kappa \left[ N \ln N + \frac{E-N(\frac{w}{2}+v)}{2v} \ln \left(-\frac{E-N(\frac{w}{2}+v)}{2v}\right) - \frac{E-N(\frac{w}{2}-v)}{2v} \ln \left(\frac{E-N(\frac{w}{2}-v)}{2v}\right) \right] \tag{32}$$

The temperature of the system is:

$$\frac{1}{T} = \left(\frac{\partial S}{\partial E}\right)_N = -\frac{\kappa}{2v} \ln \left(-\frac{E-N(\frac{w}{2}-v)}{E-N(\frac{w}{2}+v)}\right) \tag{33}$$

**Physical explanation.**

It can be seen from Eq.(33) that, if $E < \frac{N}{2}w$ then T >0, However if $E > \frac{N}{2}w$, then T <0. Now let us consider as well the variation of the entropy S with the energy E. it is observed that for $E = N\left(\frac{w}{2} - v\right)$, both S and T vanish. As E increases, they too increase until we reach the special situation where $E = \frac{N}{2}w$. The entropy is then seen to have attained its maximum value S =Nk ln 2 , while the temperature has reached an infinite value. Throughout this range, the entropy was a monotonically increasing function of energy, so T was positive. Now as E equals $[\frac{N}{2}w]_+$ , $\left(\frac{ds}{dE}\right)$ becomes $0_-$ and T becomes $-\infty$. With a further increase in the value of E, the entropy monotonically decreases; as a result, the temperature continues to be negative, though its magnitude steading decreases. Finally, we reach the largest value of E, namely $N\left(\frac{w}{2} + v\right)$ where the entropy is once again zero and $T = -0$. In figures (1) and (2) we have illustrated the thermal behaviour of the entropy S and the internal enegy U for some values of $a = \frac{v}{w}$.

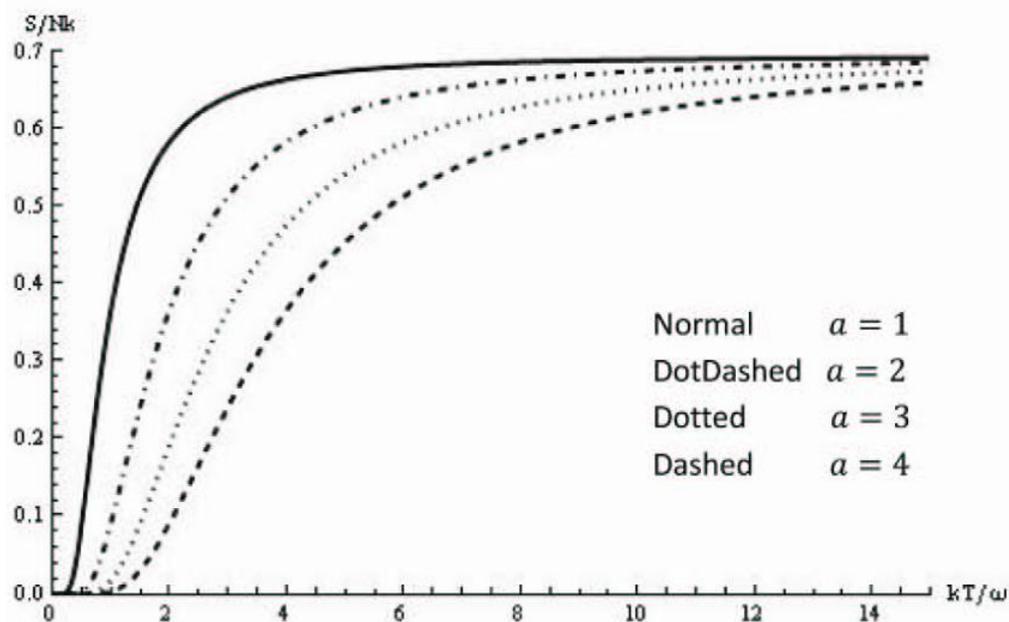

Figure 1. The entropy of a spin one half system in a constant quasianti-Hermitian quaternionic potential as a function of temperature for some values of "$a$".

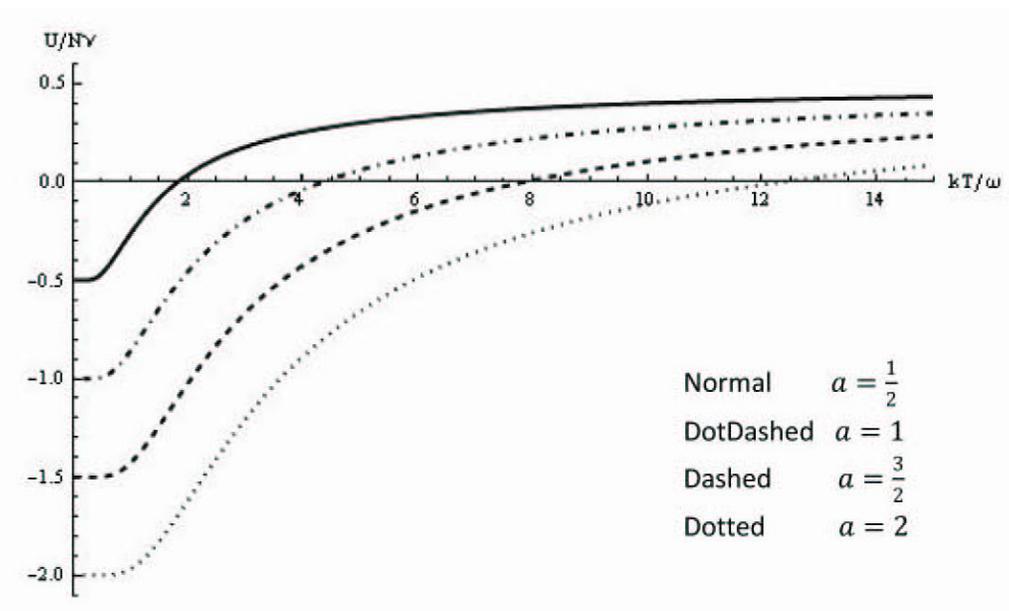

Fig. 2.

Figure 2. The internal energy of a spin one half system in a constant quasianti-Hermitian quaternionic potential as a function of temperature for some values of "$a$".

As mentioned earlier the physical system is a spin half particle in the quarternionic potential. The experimental aspects of quaternionic potentials have been investigated by Peres, McKellar, Brumby and others [8,9,10,15,16].

This system is the quasianti-Hermitian quaternioinc analogue of the ordinary(Hermitian) magnetic system i.e., a system composed of spin half particles in a constant magnetic field which is a famous system showing negative temperature discussed in the text books, see e.g., [17].

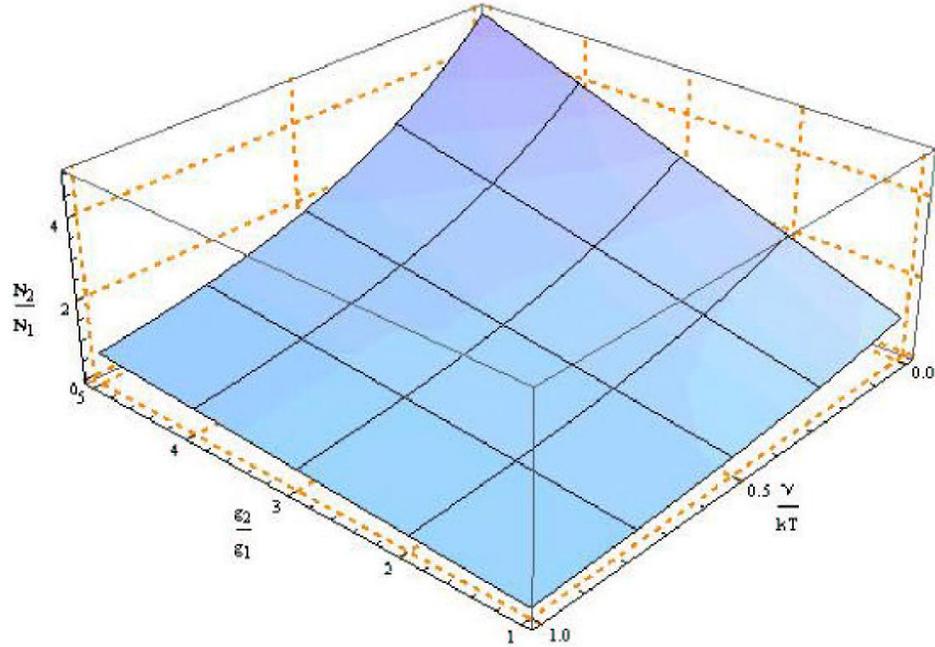

Fig. 3.

Figure 3. Population inversion versus $\frac{v}{kT}$, for some values of degeneracy.

Atomic and molecular physics are two of the oldest branches of quantum physics and they have played a major role in modern physics. One of the key and important concepts in quantum physics is the population inversion which makes for instance lasers operate. So we study the implication of negative temperature and the population inversion of the system discussed in this section in more detail. From Eq.(33), we have :

$$\frac{N_+}{N_-} = e^{-\frac{2v}{\kappa T}} \qquad (34)$$

where $v = E_+ - E_- = \Delta E$, is the energy difference between two levels. In the case of having degeneracies, Eq.(34) is replaced by :

$$\frac{N_+}{N_-} = \frac{g_2}{g_1} e^{-\frac{2v}{\kappa T}} \qquad (35)$$

where $g_2$ and $g_1$ are the degeneracies of the levels 2 and 1 respectively. From Eq.(34), we see that for $T > 0$, we have $N_- > N_+$ which means that the population of the lower level is greater than the population of the upper level, so there is more upward transitions due to absorption than downward transitions due to stimulated emission. To amplify the beam, the rate of stimulated emission transitions must exceed the rate of absorption. This implies that $N_+$ must exceed $N_-$, which is called population inversion.

From Eq.(34), we observe that population inversion corresponds to negative temperatures. It is worth to mention that in the case of having degeneracies i.e. $\frac{g_2}{g_1} > 1$ in Eq.(35), the negative temperature is not necessary to have population inversion and one can have the population inversion via the ratio $\frac{g_2}{g_1}$. The illustration of Eq.(35) versus $\frac{v}{kT}$ for some values of $\frac{g_2}{g_1}$ has been shown in Fig.(3).

## 7- Conclusion.

The notion of quaternions put forward for the first time by Hamilton. Maxwell used quaternions to formulate the equations of electrodynamics. Since then, a lot of works have been devoted to applying of quaternions to different theories of physics. On the other hand non-Hermitian Hamiltonian operators have been the subject of intense research activity in recent years. In this paper thermodynamics of quasianti-Hermitian quaternionic systems in equilibrium in the framework of canonical ensemble is studied and all the physically relevant quantities are derived. The negative temperature for such systems is also studied and a physical example is discussed.

## 8 - Acknowledgment.

I would like to thank A. Tatar, S. M. Fahimi and S. Abbaspour for their kind assistance.

## 9 - References.